\documentclass[useAMS,usegraphicx,usenatbib]{mn2e}
\DeclareMathSymbol{\not}{\mathrel}{symbols}{"36}

\title[The chemical evolution of the Galactic halo]{Are ancient dwarf satellites  the building blocks of the Galactic halo?}
\author[Spitoni et al.]{E. Spitoni$^1$\thanks{E-mail: spitoni@oats.inaf.it}, F. Vincenzo$^{1,2}$, F. Matteucci$^{1, 2, 3}$, D. Romano$^{4}$  \\
  $^1$ Dipartimento di Fisica, Sezione di Astronomia, Universit\`a di Trieste, via G.B. Tiepolo 11, I-34131, Trieste, Italy \\  
 $^2$ I.N.A.F. Osservatorio
  Astronomico di Trieste, via G.B. Tiepolo 11, I-34131, Trieste,
  Italy\\
 $^3$ I.N.F.N. Sezione di Trieste, via Valerio 2, 34134 Trieste, Italy\\
$^4$ I.N.A.F. Osservatorio
  Astronomico di Bologna, via Ranzani 1, I-40127, Bologna,
  Italy
}

\usepackage[percent]{overpic}

\begin{document}
\date{Accepted 2016 February 29. Received 2016 February 29; in original form 2015 December 10}

\pagerange{\pageref{firstpage}--\pageref{lastpage}} \pubyear{xxxx}

\maketitle

\label{firstpage}

\begin{abstract}
 According to the current cosmological cold dark matter paradigm, the
Galactic halo could have been the result of the assemblage of smaller
structures.  Here we explore the hypothesis that the classical and
ultra-faint dwarf spheroidal satellites of the Milky Way have been the
building blocks of the Galactic halo by comparing their [$\alpha$/Fe]
and [Ba/Fe] versus [Fe/H] patterns with the ones observed in Galactic
halo stars.  The $\alpha$ elements deviate substantially from the
observed abundances in the Galactic halo stars for [Fe/H] values
larger than -2 dex, while they overlap for lower metallicities. On the
other hand, for the [Ba/Fe] ratio the discrepancy is extended at all
[Fe/H] values, suggesting that the majority of stars in the halo are
likely to have been formed in situ.  Therefore, we suggest that
[Ba/Fe] ratios are a better diagnostic than [$\alpha$/Fe] ratios.
Moreover, for the first time we consider the effects of an enriched
infall of gas with the same chemical abundances as the matter ejected
and/or stripped from dwarf satellites of the Milky Way on the chemical
evolution of the Galactic halo. We find that the resulting chemical
abundances of the halo stars depend on the assumed infall time scale,
and the presence of a threshold in the gas for star formation.  In
particular,  in models with an  infall timescale for the halo around 0.8 Gyr coupled with a threshold in the surface gas density
for the star formation (4
$\mathrm{M}_{\odot}\,\mathrm{pc}^{-2}$), and the enriched infall from dwarf spheroidal satellites, the first halo stars  formed show 
      [Fe/H]$>$-2.4 dex. In this case, to explain [$\alpha$/Fe] data for stars with [Fe/H]$<$-2.4 dex we need stars formed in dSph systems.

\end{abstract}

\begin{keywords}
Galaxy: abundances - Galaxy: halo - Galaxy: evolution - ISM: abundances
\end{keywords}

\section{Introduction}
 The formation and evolution of the Milky Way (MW) halo has been the
subject of several investigations in the past years, and a great deal of 
observational work has been done in
order to obtain more and more precise abundance determinations in 
stars in the Galaxy and Local Group galaxies. We recall here a number of 
on-going and planned spectroscopic MW survey, such 
as RAVE (Steinmetz et al. 2006), SEGUE (Yanny et al. 2009),
APOGEE ( Majewski et al. 2010),
HERMES (Freeman 2010), Gaia-ESO (Gilmore et al. 2012).

A crucial information regarding the dominant mechanisms responsible
for the formation of the MW halo is encoded in the chemical and
kinematical properties of its member stars. In particular, the study
of the MW stellar halo provides several clues about the earliest
phases of the Galaxy evolution, since  the halo is the easiest
place where to find the most metal-poor and oldest stars currently
known in the Universe  although very old stars could be found also in the bulge of the MW (White \& Springel 2000).

The current cosmological $\Lambda$ cold dark matter ($\Lambda$CDM)
paradigm envisages the assemblage of large structures in the Universe
as starting from the coalescence of smaller ones, via cooling and
condensation of gas in always larger dark matter halos (Press \&
Schechter 1974; White \& Rees 1978; Springel, Frenk \& White
2006). According to the $\Lambda$CDM model, a MW-like galaxy must have
formed by the coalescence of a large number of smaller systems which,
even today, might be still in the process of being accreted. In
particular, dwarf spheroidal galaxies (dSphs) were proposed in the
past as the best candidate small progenitor systems, which merged
through cosmic time to eventually form the stellar halo component of
the Galaxy (e.g. Grebel 2005). The MW dSph satellites have been soon
recognized among the faintest and most dark matter dominated stellar
systems ever observed in the Universe, before the discovery of
ultra-faint dwarf galaxies.  Nevertheless, the role played by dSphs in
shaping the halo of the MW still remains controversial. Major issues
are the still relatively small number of discovered Galaxy satellites
(the so-called missing satellite problem, e.g. Bullock 2010, Klypin et
al. 1999, and Moore et al. 1999) and the different chemical abundance
patterns in halo and dSph stars (Shetrone et al. 2001, Venn et
al. 2004, Vincenzo et al. 2014).

In order to test the capability of the hierarchical galaxy formation
scenario to explain the MW halo metallicity distribution function
(MDF), Prantzos (2008) presented the results of an approximated
analytical model, where the Galaxy stellar halo was assumed to
assemble by means of successive merger events of small sub-halos with
similar physical properties as current dSphs.  Although the treatment
of the  interstellar medium (ISM) chemical evolution was very
simple, Prantzos (2008) claimed to reproduce the Galaxy halo MDF,
since his results rely on the stellar mass distribution function of
the merging sub-halo population, which the current hierarchical galaxy
formation paradigm can predict with very high accuracy.  Nevertheless,
Prantzos (2008) did not discuss any implication on the [$\alpha$/Fe]
versus [Fe/H] abundance pattern, which is one of the main issues which
hierarchical picture has to deal with, since the abundance patterns of
surviving Local Group galaxies do not match those of the stars in the
stellar halo.

 It is worth reminding that Unavane et al. (1996) and Jofre \& Weiss (2011) used the age distributions of stars in the halo and dSphs to test the origin of halo stars. 

Interestingly, by making use of a chemical evolution model within a
cosmological framework, Font et al. (2006) found that the discrepancy
in the [$\alpha$/Fe] ratios can be solved if the majority of the MW
halo formed by accreting sub-halos with mass in the range
10$^5$-10$^{8}$M$_{\odot}$, which had been disrupted very early ($>$
8-9 Gyr ago). 

 On the other hand, Fiorentino et al. (2015) using RR Lyrae stars as
tracers of the Galactic halo ancient stellar component, showed that
dSphs do not appear to be the major building-blocks of the
halo. Leading physical arguments suggest an extreme upper limit of
50\% to their contribution.

 In recent years, Willman et al. (2005) and  Belokurov et al (2006a,b, 2007)
 using the Sloan Digital Sky Survey (SDSS, York et al. 2000) were able 
 to discover an entirely new population of hitherto unknown
stellar systems: the so-called ultra faint dwarf spheroidal galaxies
(UfDs), which are characterized by extremely low luminosities, high
dark matter content, and very old and iron-poor stellar populations
 (Belokurov et al. 2006a; Norris et al (2008, 2010); Brown et al 2012).
  Furthermore, UfD systems are observed to be
completely gas-free at the present time.  The number of UfDs has
increased constantly in the last decade and completeness estimates
suggest that many more of these faint satellites are still to be
discovered in the Local Group (Tollerud et al. 2008). This fact might
place them as the survived building blocks of the Galaxy stellar halo,
dramatically lacking in the past.

The main aims of this work can be summarized as follows.  

\begin{enumerate} 
\item We test the hypothesis that dSph and UfD galaxies have been the  
building blocks of the Galactic halo, by assuming that the halo formed by 
accretion of stars belonging to these galaxies. 

\item We explore the scenario, in which the 
Galactic halo formed by accretion of chemically enriched gas
originating from dSph and UfD galaxies. 
\end{enumerate} 

In  Spitoni (2015), using the formalism described by Matteucci (2001), 
Recchi et al. (2008), and Spitoni et al. (2010) it was  shown, for the 
first time, an analytical solution for the evolution of the metallicity of a 
galaxy in
presence of ``environment'' effects.   In this work a galaxy suffers, during its
evolution, the infall of enriched gas from another evolving galactic
system, with this gas having chemical abundances variable in time.

In this work we  extend the results of Spitoni (2015) to detailed
chemical evolution models in which the Instantaneous Recycling Approximation 
(IRA) is relaxed,  for the particular case of the chemical
enrichment of the Galactic halo surrounded by dSph and UfD galaxies.

The paper is organized as follows: in Section 2 we present our
chemical evolution models for the Galaxy, dSph and UfD galaxies, in
Section 3 we describe the way in which we implement the enriched gas
infall on the chemical evolution of the Galactic halo.  The results
are presented in Section 4.  Finally, our conclusions are summarized
 in Section 5..

\section {The chemical evolution models} 

In this work
 for the first time we present a chemical evolution model where we
 assume that the Galactic halo is formed by accretion of enriched gas
 with chemical abundances identical to those of gas outflowing/stripped from dSph,
 and UfD galaxies.  In this section we will give some details related
 to the reference chemical evolution models considered in this paper
 for the MW, dSph, and UfD galaxies.  

For the star
 formation rate (SFR) all the Milky Way models  adopt the following Kennicutt
 (1998) like law:
\begin{equation}
\psi(t) \propto \nu \sigma_{g}^{k},
\label{k1}
\end{equation}
 where $\nu$ is the star formation efficiency,  $\sigma_g$ is the surface
gas density, and $k$ is the gas surface exponent, with an exponent k=1.5.
On the other hand, the SFRs of dSph and UfD galaxies are considered proportional to the volume gas density $\rho_{g}$  with an exponent k=1: 
\begin{equation}
\psi(t) \propto \nu \rho_{g}^{k}.
\label{k2}
\end{equation}

 In Romano et al. (2015) it is recalled that originally  Kennicutt (1998) law refers to surface densities.
 They show that for star-forming regions with
roughly constant scaleheights, the surface densities can be turned
into volume densities. Moreover, using the fact that the Kennicutt–
law indicates that the star formation rate is controlled by the self-gravity of the gas, it is possible to show the equivalence between k=1.5 in eq. (\ref{k1}) and k=1 in eq. (\ref{k2}).

\subsection{The Milky Way}
We will consider the following two reference chemical evolution models for the 
MW galaxy:
\begin{table*}

\label{TMW}
\begin{center}
\begin{tabular}{c c c c c c c c c c c c c}
  \hline
\multicolumn{10}{c}{{\it \normalsize The Milky Way: the solar neighborhood  model parameters}}\\
\\

 Models &Infall type &$\tau_H$& $\tau_D$&Threshold & $k$& $\nu$& IMF& $\omega$&\\

&  &[Gyr]& [Gyr]& [M$_{\odot}$pc$^{-2}$]&&[Gyr $^{-1}$] &&[Gyr$^{-1}$]&\\  
  
\hline

2IM & 2 infall & 0.8   & 7 & 4 (halo-thick disc) &1.5 & 2 (halo-thick disc)& Scalo (1986)&/\\
&&&&7 (thin disc)&& 1 (thin disc)\\
\hline

2IMW & 2 infall & 0.2   & 7 & 4 (halo-thick disc) &1.5 & 2 (halo-thick disc)& Scalo (1986)&14\\
&&&&7 (thin disc)&& 1 (thin disc)\\

\hline
\end{tabular}
\end{center}
\caption{Parameters of the chemical evolution models for the Milky Way (Brusadin et al. 2013) in the solar neighborhood.}
\end{table*}

\begin{table*}
\begin{tabular}{c c c c c c c c c c c c}
\hline
\multicolumn{12}{c}{{\it \normalsize dSphs: parameters of the model}}\\
\\

$\nu$& $k$& $\omega$ & $\tau_{inf}$ & SFH (99\% of stars) & $M_{inf}$ & $M_{DM}$ & $r_{L}$ & $S=\frac{r_{L}}{r_{DM}}$ & IMF& $t_{gw}$ & [Fe/H]$_{peak}$\\

[Gyr$^{-1}$] && & [Gyr] & [Gyr] & [$M_{\odot}$] & [$M_{\odot}$] & [pc] & & & [Gyr]&[dex]\\
\hline

0.1 & 1&10 & $0.5$ & 0-2.43 & $10^{7}$ & $3.4\cdot10^{8}$ & $260$ & $0.52$ & Salpeter (1955) & 0.013& -2.10\\
\hline
\end{tabular}
\caption{ Parameters of the chemical evolution model for a general dSph galaxy. }
\end{table*}

\begin{table*}
\begin{tabular}{c c c c c c c c c c c c}
\hline
\multicolumn{12}{c}{{\it \normalsize UfDs: parameters of the model}}\\
\\

$\nu$& $k$& $\omega$ & $\tau_{inf}$ & SFH (99\% of stars) & $M_{inf}$ & $M_{DM}$ & $r_{L}$ & $S=\frac{r_{L}}{r_{DM}}$ & IMF& $t_{gw}$&[Fe/H]$_{peak}$ \\

[Gyr$^{-1}$] && & [Gyr] & [Gyr] & [$M_{\odot}$] & [$M_{\odot}$] & [pc] & & & [Gyr]&[dex] \\
\hline

0.01 & 1&10 & $0.001$ & 0-0.49 & $10^{5}$ & $10^{6}$ & $35$ & $0.1$ & Salpeter (1955)& 0.088 &-3.30\\
\hline
\end{tabular}
\caption{Parameters of the chemical evolution model for a general dSph galaxy. }
\end{table*}

\begin{enumerate}

\item The  classical two-infall  model (2IM) presented by Brusadin  et al. (2013), which is an updated version of the  two-infall model of Chiappini et al. (1997). The Galaxy is assumed to have
formed by means of two main infall episodes: the first formed the halo
and the thick disc, the second the thin disc.  The accretion law of a
certain element $i$ at the time $t$ and Galactocentric distance $r$ is
defined as:
\begin{equation}
A(r,t,i)=X_{A_{i}}\left( a(r) e^{-t/ \tau_{H}(r)}+ b(r) e^{-(t-t_{max})/ \tau_{D}(r)} \right).
\label{a}
\end{equation}
The quantity $X_{A_{i}}=\sigma_i(t)/\sigma_{gas}(t)$ is the abundance
by mass of the element $i$ in the infalling material, while
$t_{max}=1$ Gyr is the time for the maximum infall on the thin disc,
$\tau_{H}= 0.8$ Gyr is the time scale for the formation of the halo
and thick-disc and $\tau_{D} (r)$ is the timescale for the formation
of the thin disc and is a function of the galactocentric distance
(inside-out formation, Matteucci \& Fran\c cois, 1989; Chiappini et
al. 2001). In the 2IM model the abundances $X_{A_{i}}$ show primordial
gas compositions and are constant in time.
 
Finally, the coefficients $a(r)$ and $b(r)$ are obtained by imposing a
 fit to the observed current total surface mass density in the
 different considered Galactic components as a function of the
 Galactocentric distance; for instance, for the thin disc:
\begin{equation}
\sigma(r)=\sigma_{0}e^{-r/r_{D}},
\end{equation}
where $\sigma_{0}$=531 $M_{\odot}$ pc$^{-2}$ is the central total
surface mass density and $r_{D}= 3.5$ kpc is the scale length.
The halo surface mass density at solar
position is quite uncertain and we assume it to be 17 $M_{\odot}$ pc$^{-2}$ in the solar neighborhood.
In fact,  the  total  surface  mass  density  in  the  solar  vicinity  is
$\sim$ 71 $M_{\odot}$ pc$^{-2}$ (Kuijken \& Gilmore 1991) with $\sim$ 54 $M_{\odot}$ pc$^{-2}$ corresponding to  the  disc  surface mass  density.

\item The two-infall model plus outflow of Brusadin et al. (2013; in this work we will indicate it as the 2IMW model). In this model  a gas outflow occuring
during the halo phase with a rate proportional to the star formation
rate through a free parameter is considered.  
  Following Hartwick (1976), the outflow rate is defined as:
\begin{equation}
\frac{d \sigma_w}{dt}=-\omega \psi(t),
\end{equation}
where $\omega$ is the outflow efficiency.

\end{enumerate}

In Table 1 the principal characteristics of the two chemical evolution
models for the MW are summarized: in the second column the
infall type is indicated, in the third, and in the fourth columns the
time-scale $\tau_H$ of halo formation and the time-scale $\tau_D$ of
the thin disc formation, are drawn. The adopted threshold in the
surface gas density for the star formation (SF) is reported in columns
5. In column 6 the exponent of the Schmidt (1959) law is shown, in
columns 7 and 8 we report the star formation efficiency and the IMF,
respectively. Finally, in the last column the presence of the wind is
indicated in term of the efficiency $\omega$.

\begin{figure}
\centering \includegraphics[scale=0.4]{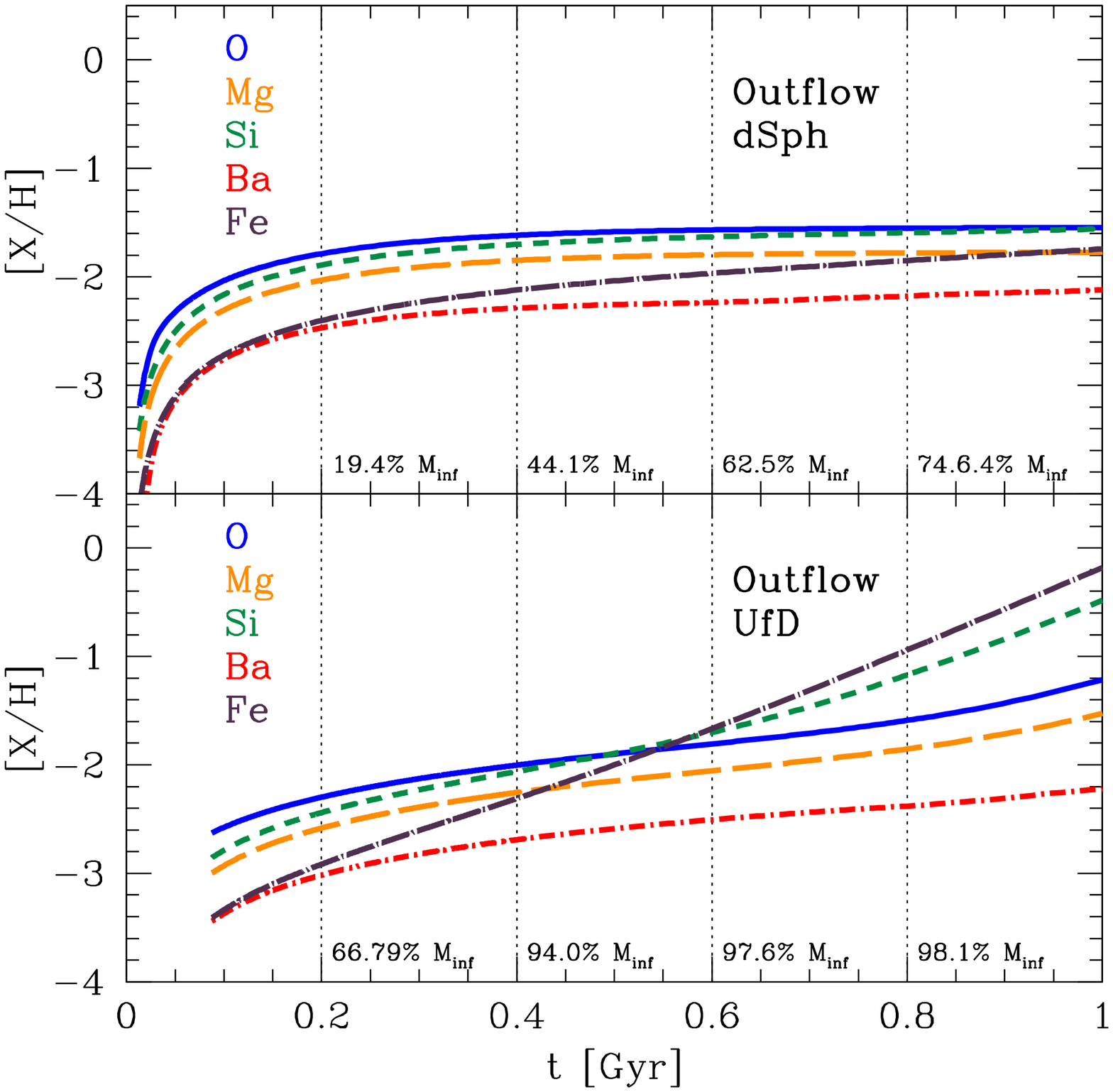} 
\centering \includegraphics[scale=0.4]{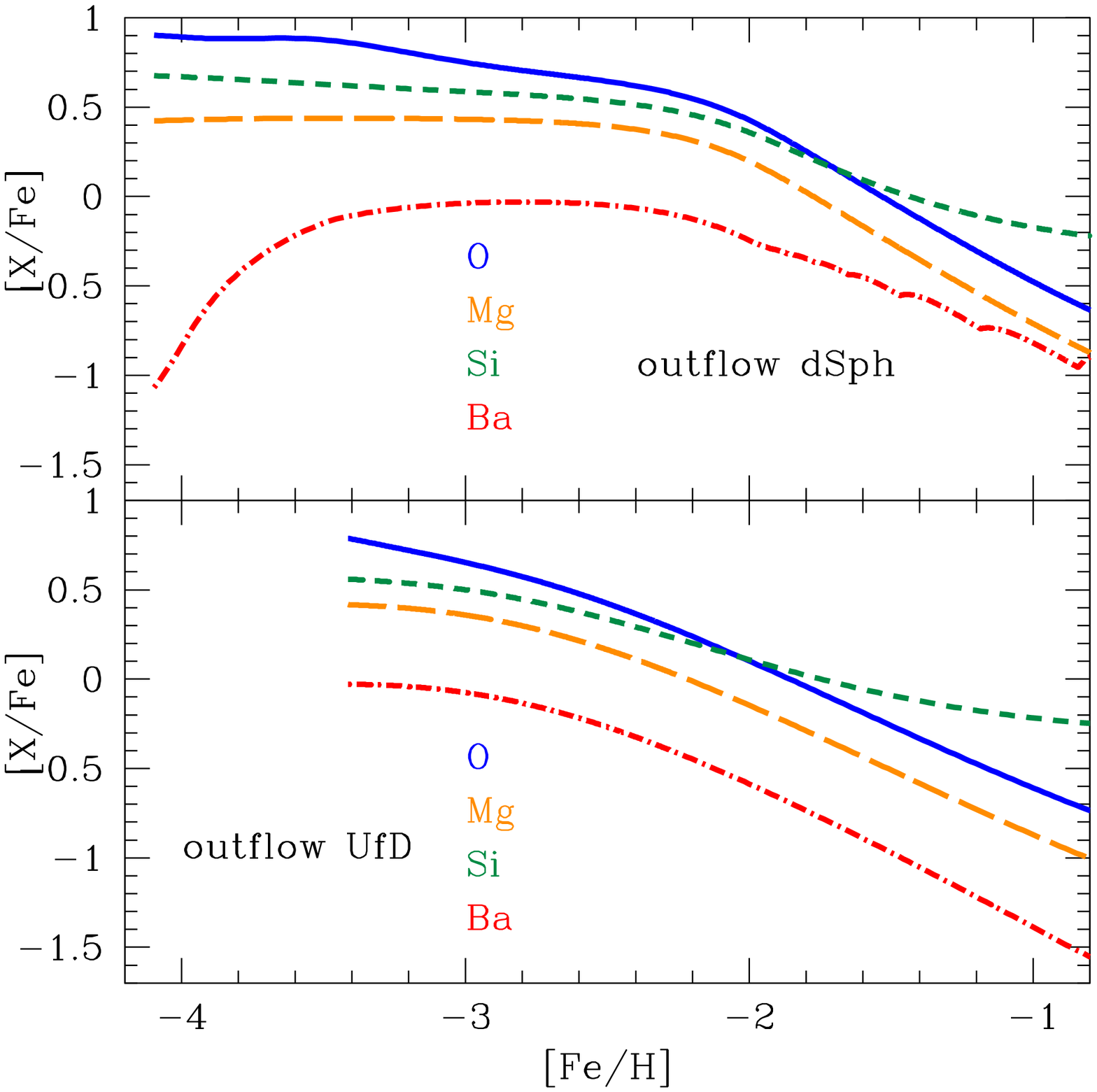}
 \caption{{\it Upper panels}:  The evolution in time of the chemical abundances for  O, Mg, Si,  Ba, Fe in  the gas ejected  as galactic wind from dSphs and UfDs. As shown in Tables 2 and 3 in UfDs the onset of the wind happens at later times compared with  dSph objects: $t_{gw}(dSph)< t_{gw}(UfD)$.  We also indicate the cumulative ejected gas mass by outflows at 0.2, 0.4, 0.6, and 0.8 Gyr in terms of percentage of the infall mass $M_{inf}$. 
{\it Lower panels}: The abundance ratio [X/Fe] as a function of [Fe/H] for the following chemical elements: O, Mg, Si, and Ba  of the outflowing gas ejected by a  dSph galaxy, and by a UfD galaxy.}
\label{wind1}
\end{figure}

\begin{figure}
	  \centering \includegraphics[scale=0.42]{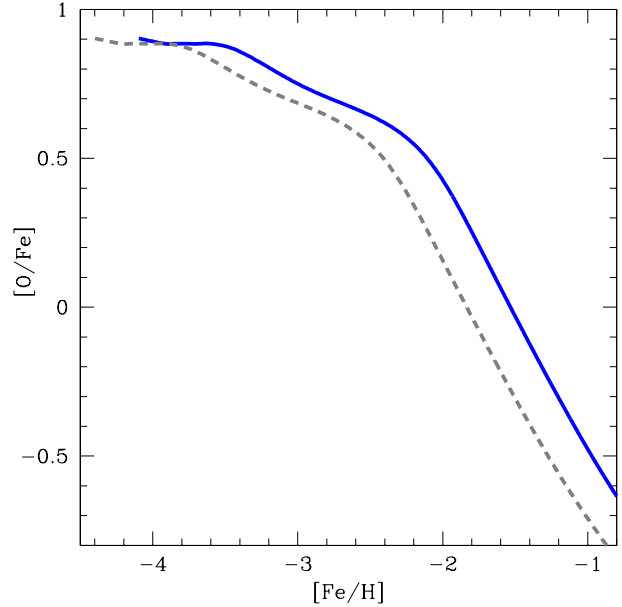} 
\caption{The abundance ratio [O/Fe] as a function of [Fe/H]   of the outflowing gas ejected by the  dSph galaxy  (blue solid line)  compared  with the ``mixed'' infall model ( dashed gray line).  }
\label{wind2}
\end{figure} 

In this work we only focus on the study of the halo phase, investigating the
effects of the pre-enriched infall of gas assuming chemical abundances
taken from the outflowing/stripped gas from dSph and UfD galaxies. We assume
that the halo phase spans the range of [Fe/H] up to -1 dex.  It is
important to underline that in our model we do not modify the gas
infall laws of the Brusadin et al. (2013) model, and the way in which the
Galaxy is built up remains the same. {\it Here, we only consider a
time dependent enriched infall, i.e. $X_{A_{i}}(t)$, with the same
chemical abundances of the outflowing gas from dSph and UfD galaxies.}

Concerning the observational data for the $\alpha$ elements and Fe, as
done in Micali et al. (2013), we employ only data in which NLTE
corrections are considered. In particular, the data we use for 
Galactic Halo stars are from Gratton et al.  (2003), Cayrel et al.  (2004),
Akerman et al. (2004), Mashonkina et al. (2007) and Shi et al. (2009).
For  Ba we use the data of Frebel  (2010), as selected and binned by Cescutti et al. (2013).

\begin{figure}
\centering	  \includegraphics[scale=0.42]{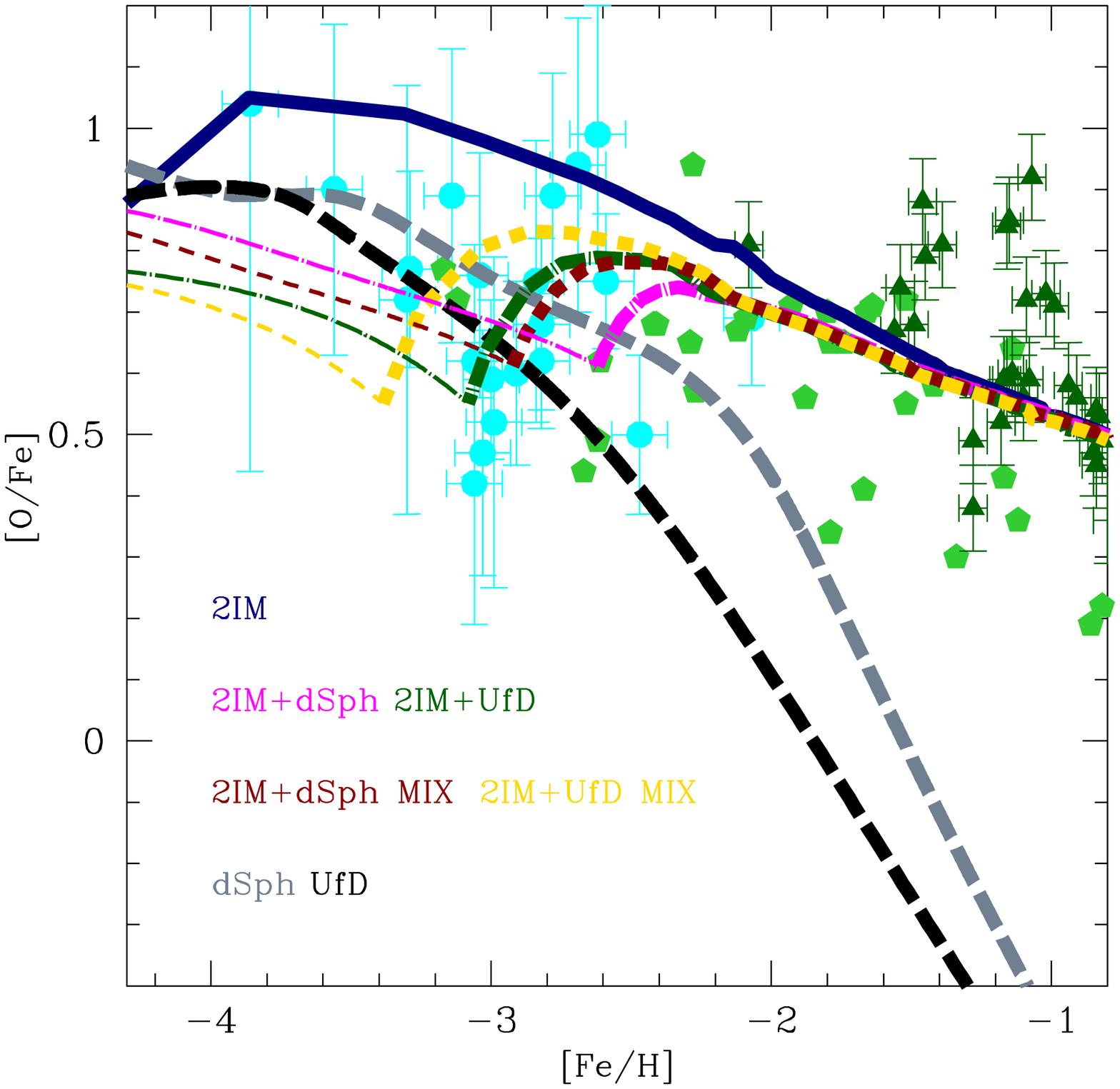}
\caption{The abundance ratio [O/Fe] as a function of [Fe/H] in the solar
neighborhood for the reference model 2IM is drawn with the solid blue
line. {\it Models of the Galactic Halo with the enriched infall from
dSph}: the magenta dashed dotted line and the red short dashed line
represent the models 2IM+dSph and 2IM+dSph MIX, respectively. {\it
Models of the Galactic Halo with the enriched infall from UfDs}: the
green dashed dotted line and the yellow short dashed line represent
the models 2IM+UfD and 2IM+UfD MIX, respectively.   Thinner lines
indicate the ISM chemical evolution phases in which the SFR did not
start yet in the Galactic halo, and during which stars are no
created.  {\it Models of the dSph and UfD galaxies}: The long dashed
gray line represents the abundance ratios for the dSph galaxies,
whereas long dashed black line for the UfD galaxies. {\it
Observational data of the Galactic Halo:} Cayrel et al. (2004) (cyan
circles), Akerman et al.  (2004) (light green pentagons), Gratton et
al. (2003) (dark green triangles).  }
\label{O1}
\end{figure}

\subsection{The dSph and UfD  galaxies}

 \begin{figure}
	  \centering \includegraphics[scale=0.42]{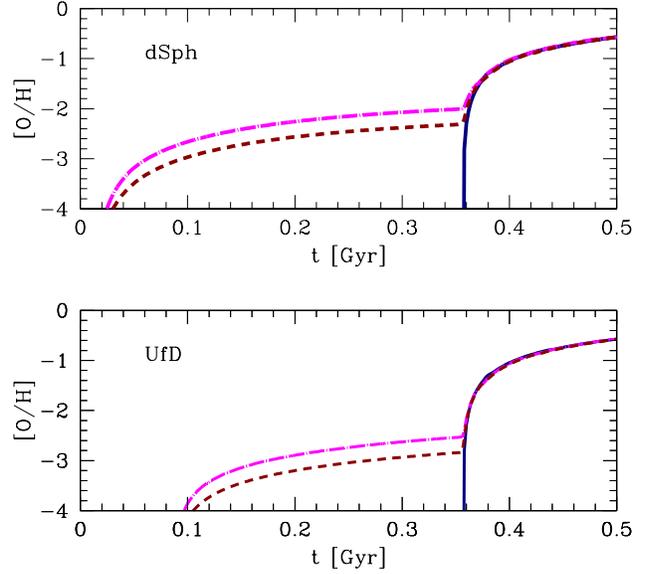}
\caption{ {\it Upper panel}: The abundance ratio [O/H] as a function of Galactic time  in the solar
neighborhood. We compared the 2IM model (blue solid line) with the
model where we have taken into account the enriched infall from dSph
galaxies  (model 2IM+dSph with  dashed dotted magenta line). With the short dashed
red line we represent the model 2IM+dSph MIX. {\it Lower panel}: As in the upper panel but considering the enriched gas from UfD galaxies.}
\label{OHt}
\end{figure}
To model the chemical evolution of dSph and UfD galaxies we refer to
the work of Vincenzo et al. (2014).  In Tables 2 and 3 the main
parameters of generic models for ``classical'' dSph and UfD galaxies
are reported, respectively. The star formation efficiency $\nu$, the
exponent $k$ of the Kennicutt (1998) law, and the wind efficiency
$\omega$ are drawn in column one, two and three, respectively. In the
other columns are reported: the infall timescale (column 4), the
period of major star formation activity  in which the 99\% of
stars are formed (column 5); total infall gas mass (column 6); mass
of the dark matter halo (column 7); effective radius of the luminous
(baryonic) matter (column 8); ratio between the core radius of the DM
halo and the effective radius of the luminous matter (column 9); 
in column 10 the adopted IMF is indicated.  In column 11 the time of
the onset of the galactic wind is reported. In the last column we show
the [Fe/H] abundance of the peak of the predicted G-dwarf metallicity
distribution.  We assume that UfD objects are characterized by a very
small star formation efficiency (0.01 Gyr$^{-1}$) and by an extremely
short time scale of formation (0.001 Gyr).  Hence UfD objects started
to form stars as most of their infall mass,
$M_\mathrm{inf}=10^{5}\,\mathrm{M}_{\sun}$, was accumulated in their
DM potential well ($\tau_{\mathrm{inf}}=0.001\,\mathrm{Gyr}$) and the
star formation efficiency is very low
($\nu=0.01\,\mathrm{Gyr}^{-1}$). Our chemical evolution model for a
typical dSph galaxy assumes an infall mass
$M_\mathrm{inf}=10^{7}\,\mathrm{M}_{\sun}$, an infall timescale
$\tau_\mathrm{inf}=0.5\,\mathrm{Gyr}$, and a star formation efficiency
$\nu=0.1\,\mathrm{Gyr}^{-1}$.  {\it Although the star formation
history (SFH) is assumed to be extended over the entire galaxy
lifetime} both for the dSph and the UfD galaxy, it is strongly
concentrated in the earliest stages of the galaxy evolution; in fact,
as most of the infall mass has been accumulated and the galactic wind
has started, the intensity of the SFR becomes negligible. We
acknowledge that this kind of SFR history is not representative in
particular of the more luminous dSph galaxies (e.g. Weisz et al 2014).
  We point out that in the modeling the dSphs and UfDs we did not
consider any threshold in the gas density for star formation, as in Vincenzo et al. (2014). 

The time
at which the galactic winds start in  dSph and UfD
models. For the dSph the galactic wind occurs at 0.013 Gyr after the
galactic formation, whereas for UfDs at 0.088 Gyr.

As expected, the  UfD galaxies develop a  wind at later times  because of the 
smaller adopted star formation efficiency (SFE).
  Moreover, because of the shorter formation time-scale the UfDs show the G-dwarf metallicity peak at lower [Fe/H] values compared to dSph galaxies.

\begin{figure}
	  \centering \includegraphics[scale=0.42]{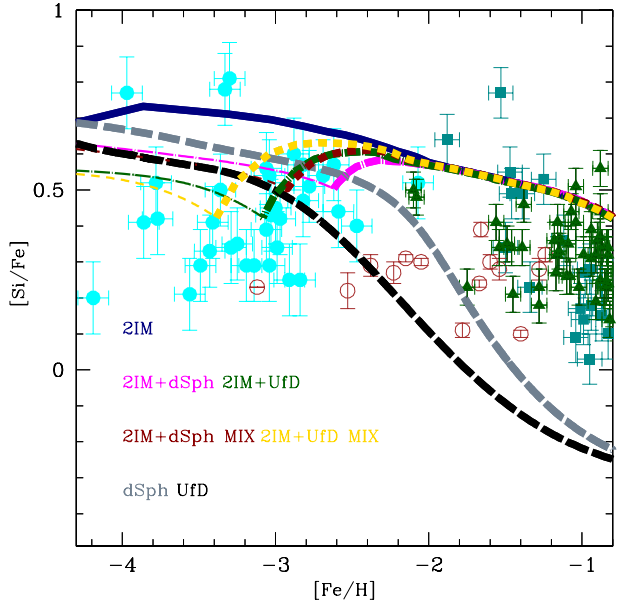} 
 \caption{ The abundance ratio [Si/Fe] as a function of [Fe/H] in the solar
neighborhood. {\it Models of the Galactic Halo with the enriched infall from dSph}:  the magenta dashed dotted line and the red short dashed line represent the models 2IM+dSph and 2IM+dSph MIX, respectively. {\it Models of the Galactic Halo with the enriched infall from UfDs}:   the green dashed dotted line and the yellow short dashed line represent the models 2IM+UfD and 2IM+UfD MIX, respectively. {\it Models of the dSph and UfD galaxies}: The long dashed gray line represents the abundance ratios for the  dSph galaxies, whereas long dashed black line for the UfD galaxies.  {\it Observational data of the Galactic Halo:} Cayrel et al. (2004)  (cyan circles), Shi et al. (2009) (open brown circles), Reddy et al. (2006) (filled blues squares),  Gratton et al. (2003) (dark green triangles). } 
\label{Si1}
\end{figure}

\subsection{Nucleosynthesis prescriptions}

 In this work, we adopt the nucleosynthesis prescriptions of Romano et al. (2010, model 15), who provide a compilation of stellar yields able to reproduce several chemical abundance patterns in the solar neighborhood. 
In particular, they assume the following sets of stellar yields. 
\begin{enumerate}
\item  For low- and intermediate-mass stars (0.8-8 M$\odot$), they include the metallicity-dependent stellar yields of Karakas (2010). 
For SNe Ia, the adopted nucleosynthesis prescriptions are from Iwamoto et al. (1999). 
\item  For massive stars (M$>$8$_{\odot}$), which are the progenitors of either SNe II or HNe, depending on the explosion energy, they assume the metallicity-dependent  He, C, N and O stellar yields, 
as computed with the Geneva  stellar evolutionary code, which takes into account the combined effect of mass loss and rotation  (Meynet \&  Maeder 2002, Hirschi  et  al.  2005, Hirschi  2007, Ekstr\"om et al. 2008); 
for all the elements heavier than oxygen, they assume the up-to-date stellar evolution calculations by 
Kobayashi et al. (2006). 

\end{enumerate} 

For  barium, we assume the stellar yields of Cescutti et al. (2006, model 1, table 4). 
In particular, Cescutti et al. (2006) includes the metallicity-dependent stellar yields of Ba as computed by Busso et al. (2001), in which barium is produced by low-mass AGB stars, 
with mass in the range $1.0\le M \le3.0$ M$_{\sun}$, as an s-process neutron capture element. A second channel for the Ba-production was 
included by Cescutti et al. (2006), by assuming that massive stars in their final explosive stage are capable of synthesizing Ba  as a primary 
r-process element. Such r-process Ba producers have mass in the range $12\le M \le30$ M$_{\sun}$.  

We remark on the fact that the contribution to barium from massive
stars was empirically computed by Cescutti et al. (2006), by matching
the [Ba/Fe] versus [Fe/H] abundance pattern as observed in the Galactic
halo stars.  They assumed for massive stars the iron stellar yields of
Woosley \& Weaver (1995), as corrected by Fran\c cois et al. (2004).

\begin{figure}
	  \centering \includegraphics[scale=0.42]{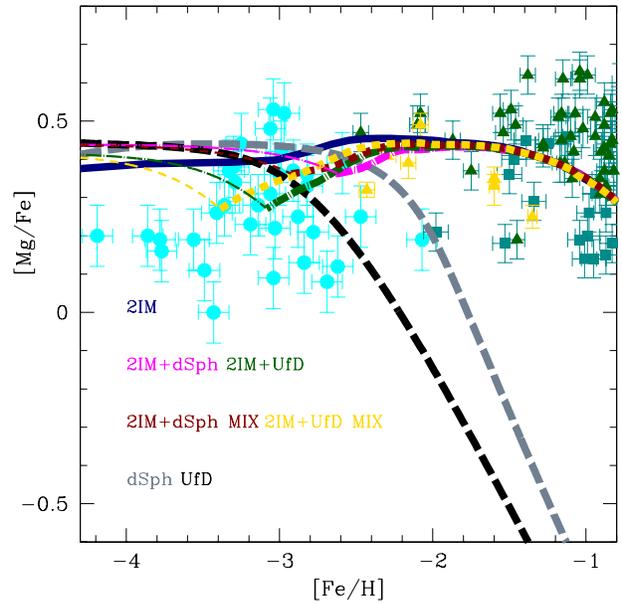} 
 \caption{ The abundance ratio [Mg/Fe] as a function of [Fe/H] in the solar
neighborhood. {\it Models of the Galactic Halo with the enriched infall from dSph}:  the magenta dashed dotted line and the red short dashed line represent the models 2IM+dSph and 2IM+dSph MIX, respectively. {\it Models of the Galactic Halo with the enriched infall from UfDs}:   the green dashed dotted line and the yellow short dashed line represent the models 2IM+UfD and 2IM+UfD MIX, respectively. {\it Models of the dSph and UfD galaxies}: The long dashed gray line represents the abundance ratios for the  dSph galaxies, whereas long dashed black line for the UfD galaxies. {\it Observational data of the Galactic Halo:} Cayrel et al. (2004)  (cyan circles), Moshonkina et al. (2007) (yellow triangles), Reddy et al. (2006) (filled blues squares),  Gratton et al. (2003) (dark green triangles). }
\label{Mg1}
\end{figure}

\begin{figure}
	  \centering \includegraphics[scale=0.42]{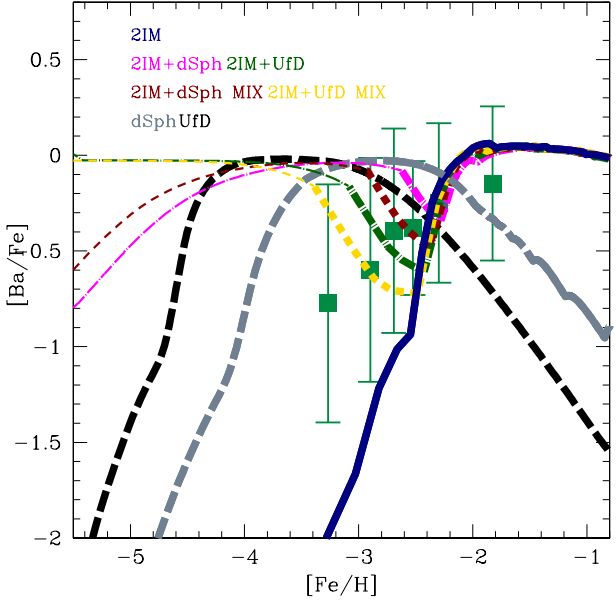}
      
\caption{The abundance ratio [Ba/Fe] as a function of [Fe/H] in the solar
neighborhood.  {\it Models of the Galactic Halo with the enriched infall from dSph}:  the magenta dashed dotted line and the red short dashed line represent the models 2IM+dSph and 2IM+dSph MIX, respectively. {\it Models of the Galactic Halo with the enriched infall from UfDs}:   the green dashed dotted line and the yellow short dashed line represent the models 2IM+UfD and 2IM+UfD MIX, respectively. {\it Models of the dSph and UfD galaxies}: The long dashed gray line represents the abundance ratios for the  dSph galaxies, whereas long dashed black line for the UfD galaxies. {\it Observational data of the Galactic Halo:} Frebel  (2010). } 
\label{Ba1}
\end{figure}

\section{The enriched infall of gas}

The novelty of this work is to take into account in a self-consistent
way time dependent abundances, i.e. $X_{A_{i}}=X_{A_{i}}(t)$ instead
of $X_{A_{i}}$ = constant in equation (\ref{a}) for the accreting gas
in the halo phase, with the values of $X_{A_{i}}(t)$ corresponding to
the chemical abundances of the material ejected from dSph and UfD
galaxies by means of their galactic winds.  Actually, it may well be
that the gas heated by SN explosions is stored in a hot gaseous halo
surrounding the satellites, from which it is stripped 
owing to the interaction with the our Galaxy. In the following, we
will only mention galactic winds for the sake of simplicity, but this
alternate option (stripping) would work equally well.
The gas infall law is the same as in the 2IM or 2IMW models and we only consider a time dependent chemical composition of the 
infall gas mass.

We take into account the enriched infall from dSph and UfD galaxies predicted by the following 2 models:

\begin{itemize}

\item Model i): The infall of gas which forms the Galactic halo is considered primordial up to the time at which the galactic wind  in dSphs (or UfDs) starts.  After this moment, the infalling gas presents the chemical abundances of the  wind. It is important to underline that for all the Galactic models  the gas infall laws are identical to the reference model presented in Brusadin et al. (2013). As stressed in the  previous section,  the only thing we are modifying here is the chemical composition of the infalling gas.  In Figs.  we refer to this model with the label ``Name of the reference model+dSph''or ``Name of the reference model+UfD''.

\item Model ii): we explore the case of a diluted infall of gas during the MW halo phase. In particular, after the galactic wind develops in the dSph (or UfD) galaxy,  
the infalling gas has a chemical composition which, by $50$ per cent,
is contributed by the dSph (or UfD) outflows; the remaining $50$ per
cent is contributed by primordial gas of a different extra-galactic
origin (in agreement with the work of Fiorentino et al. 2015). As
stated above, the infall law follows the one assumed in the 2IM or
2IMW models presented in Brusadin et al. (2013). In all the successive
figures and in the text, we refer to these models with the labels
``Name of the MW model+dSph (or UfD) MIX''.  

\end{itemize}

 In the two upper panels of Fig. \ref{wind1}, we show the
evolution in time of the chemical composition of the outflowing gas
from the dSph and the UfD galaxy for O, Mg, Si, Ba and Fe. It is worth noting that in the outflows from UfD galaxies the Fe and Si abundances are larger than in the outflows from dSphs.

 We recall that Fe is mostly produced by Type Ia SNe and Si is also produced in a non negligible amount by the same SNe. Because in our models the ratio between the time scale of formation between UfD and dSph is extremy low ($\tau_{inf}$(UfD)/$\tau_{inf}$(dSph)=$2 \times 10^{-3}$, at later times the pollution from Type IA SN  is more evident in the UfD outflow.
As shown in Tables 2 and 3, in UfDs the onset of the wind happens at later times compared with  dSph objects: $t_{gw}(dSph)< t_{gw}(UfD)$.  

 In the two
lower panels  the [$X$/Fe] versus [Fe/H] abundance patterns are presented,
where $X$ corresponds to O, Mg, Si, and Ba.

 In
Fig. \ref{wind2} is shown the [O/Fe] ratio as a function of [Fe/H] in the galactic
wind of a classical dSph galaxy with characteristics described in 
Table 2, and compared with the ``MIX'' case where the abundances are
diluted by $50$ per cent by gas of primordial chemical
composition. As stated above, in our
chemical evolution model for the dSph the galactic wind begins at
0.013 Gyr after the galaxy formation, and at 0.088 Gyr for the UfD
galaxies.

\section{The Results}

In this section, we present the results of our chemical evolution
models for the Galactic halo, by assuming that either 

\begin{itemize}
\item  A) all the stars
of the Galactic halo were born in situ in dSph galaxies

\item   B) the
Galactic halo formed by accretion of pre-enriched material,
originating in dSphs and UfDs  (Model i). We explore also the case in which the
infalling enriched material is diluted by pristine gas of different
extra-galactic origin  (Model ii). 
\end{itemize}

We divide the results in two  separate
subsections, according to the MW chemical evolution model which is
assumed, namely the 2IM (two-infall) or the 2IMW (two infall plus
outflow) models.
 It is important to stress out that the models follow the
chemical abundances in the ISM and we compare our predictions with
stellar abundances under the assumption that their atmospheres reflect
the abundances of the ISM out of which they formed. 
\subsection{The Results: the Galactic halo in the model 2IM}

In Fig. \ref{O1}, the predicted [O/Fe] versus [Fe/H] abundance
patterns are compared with the observed data in Galactic halo
stars.  In order to
directly test the hypothesis that Galactic halo stars have been
stripped from dSph or UfD systems, we show the predictions of chemical
evolution models for a typical dSph and UfD galaxy (long dashed lines
in grey and black, respectively).  The two models cannot explain the
[$\alpha$/Fe] plateau which Galactic halo stars exhibit for
$\mathrm{[Fe/H]}\ga-2.0$ dex; in fact, halo stars have always larger
[O/Fe] ratios than dSph and UfD stars.
 
\begin{figure}
	  \centering \includegraphics[scale=0.42]{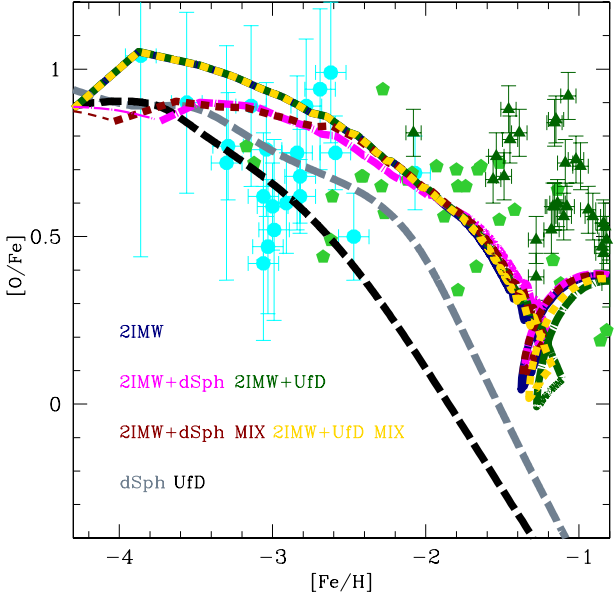}
\caption{The abundance ratio [O/Fe] as a function of [Fe/H] in the solar
neighborhood as in Fig. \ref{O1} but for the 2IMW model. } 
\label{OW}
\end{figure}

Moreover, in Fig. \ref{O1} we show the effects of the enriched infall
with chemical abundances taken by the outflowing gas from dSph and Ufd
objects on the [O/Fe] versus [Fe/H] relation. In particular, we compare
the reference 2IM model for the Galactic Halo (which assumes
primordial infall) with the ``2IM+dSph'' and ``2IM+UfD'' models
(enriched infall following Model i) and ii) prescriptions,
respectively); in the same figure, we show also the ``2IM+ dSph MIX''
and ``2IM+UfD MIX'' models, with chemical abundances of the outflowing gas
from dSph and UfD being diluted with primordial ones.

First we analyze the results with the enriched infall coming from dSph
galaxies.  We see that for oxygen we obtain a better agreement with
the data in the halo phase when we consider the enriched infall
models.  We recall that a key ingredient of the 2IM model is the
presence of a threshold in the gas density in the star formation (SF)
fixed at 4 M$_{\odot}$ pc$^{-2}$ in the Halo-thick disc phase.  During
the halo phase such a critical threshold is reached only at
$t=0.356\,\mathrm{Gyr}$ from the Galaxy formation.  On the other hand,
when including the environmental effect, we have to consider also the
time for the onset of the galactic wind, which in the dSph model
occurs at $t_\mathrm{gw}=0.013\,\mathrm{Gyr}$.

Therefore, the SF begins after 0.356 Gyr from the beginning of Galaxy
formation, and this fact explains the behavior of the curves with
enriched infalls in Fig.  \ref{O1}: during the first 0.356 Gyr in both
``2IM+dSph'' and ``2IM+dSph MIX'' models, no stars are created, and
the chemical evolution is traced by the exponential gas accretion with
a time dependent chemical enrichment (eq. \ref{a}).

 In Fig. \ref{O1} and  in all the successive figures we indicate with  thinner lines the ISM chemical evolution phases in which the SFR did not start yet in the Galactic halo, and during which stars are no created. 
 
To summarize, for the  model ``2IM+dSph'' we  distinguish three different phases in the halo chemical evolution:
\begin{itemize}

\item Phase 1): 0-0.013 Gyr, the infall is primordial, the wind in dSphs has not started yet,  and there is no SF;
\item Phase 2): 0.013-0.356 Gyr, the infall is enriched by dSphs, the SFR is zero in this phase;

\item Phase 3): 0.356-1 Gyr; the infall is enriched by dSphs, the SFR is different from zero.
\end{itemize} 

During  phase 3), the SF takes over, and increases
the [O/Fe] values because of the pollution from massive stars on short
time-scales.

\begin{figure}
      \includegraphics[scale=0.42]{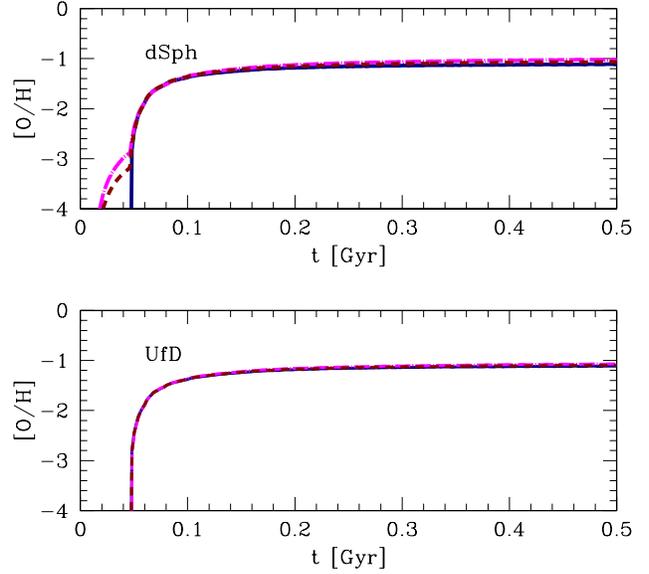}
\caption{As in Fig. \ref{OHt} the evolution in time for oxygen  but for the 2IMW model.}
\label{OHtW}
\end{figure}

 We note that the entire spread of the data cannot be explained
 assuming a time dependent enriched infall with the same abundances
 of the outflowing gas from dSph galaxies, even if there is a
 better agreement with the halo data in comparison to the model with
 primordial infall.

It is important to underline that, until the SF is  non-zero, no stars are
created; however, since our models follow the chemical abundances in
the gas phase, the solely contribution to the ISM chemical evolution
before SF begins is due to the time dependent enriched infall. It means that
in the ``2IM+dSph'' model the first stars that are formed have [Fe/H]
values larger than -2.4 dex.

In this case, to explain data for stars with  [Fe/H] smaller than -2.4 dex
we need stars formed in dSph systems (see the model curve of the
chemical evolution of dSph galaxies).
\begin{figure}
	  \centering \includegraphics[scale=0.42]{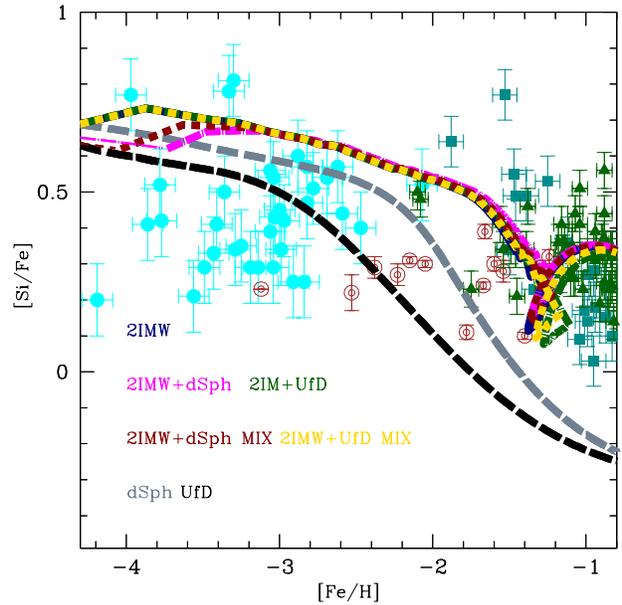} 
\caption{The abundance ratio [Si/Fe] as a function of [Fe/H] in the solar
neighborhood. As in Fig. \ref{Si1} but for the 2IMW model.  }
\label{siw}
\end{figure}

Concerning the results with the enriched infall from UfD outflow
abundances, we recall here that in our reference model for UfD
galaxies, the wind starts at 0.08 Gyr. The model results for the halo
still reproduce the data but with the same above mentioned caveat.

In Fig. \ref{OHt} we show the time evolution of oxygen abundances for
the model 2IM (primordial infall), 2IM+dSph, 2IM+dSph MIX, 2IM+UfD,
and 2IM+UfD MIX.  We notice that the reference model 2IM, as
explained before, shows chemical evolution after t=0.356 Gyr, and the
models with enriched infalls which show the fastest chemical
enrichment are the ones with infall abundances taken from the outflows of
dSph objects, because the galactic winds occur earlier than in UfD
systems.
\begin{figure}
	  \centering \includegraphics[scale=0.42]{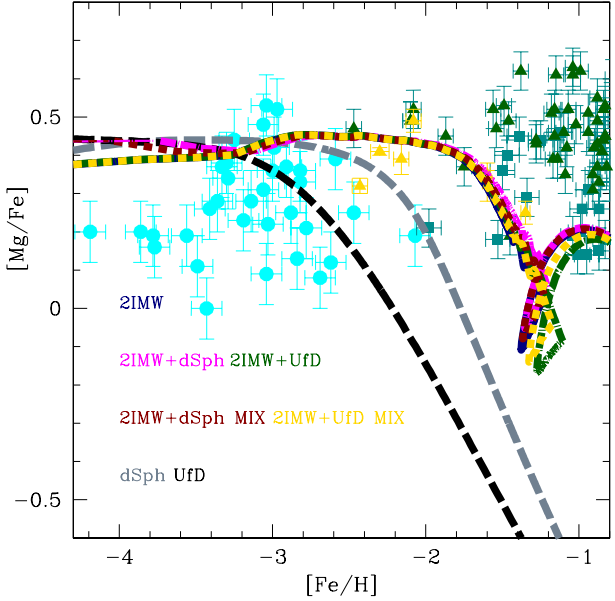} 
\caption{The abundance ratio [Mg/Fe] as a function of [Fe/H] in the solar
neighborhood. As in Fig. \ref{Mg1} but for the 2IMW model. }
\label{mgw}
\end{figure}
\begin{figure}
	  \centering \includegraphics[scale=0.42]{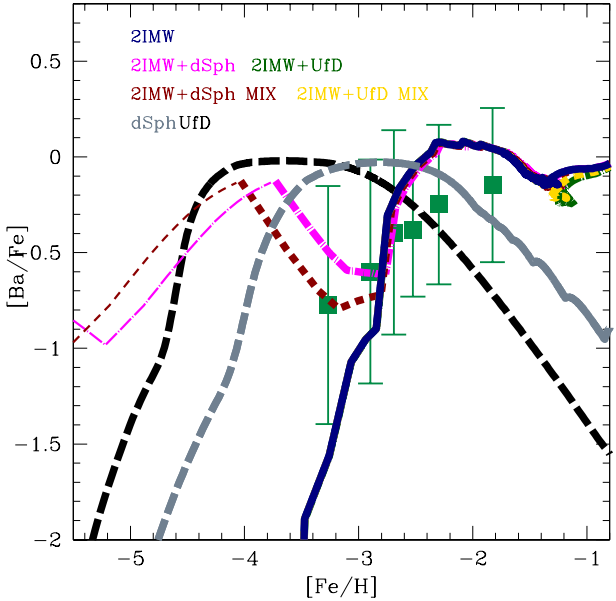} 
\caption{The abundance ratio [Ba/Fe] as a function of [Fe/H] in the solar
neighborhood. As in Fig. \ref{Ba1} but for the 2IMW model.  }
\label{BaW}
\end{figure}

In Figs. \ref{Si1} and \ref{Mg1}, we show the results of all our
chemical evolution models with the 2IM scenario for the [Si/Fe] and
[Mg/Fe] versus [Fe/H] abundance patterns, respectively.  The various
curves with different colors represent the same chemical evolution
models as in Fig. \ref{O1}. As concluded for the [O/Fe] versus [Fe/H]
abundance diagram, our reference chemical evolution models for dSph
and UfD galaxies cannot explain the observed Galactic halo data over
the entire range of [Fe/H] abundances. This rules out the hypothesis
that all Galactic halo stars were stripped or accreted in the past
from dSphs or UfDs.  This result is in agreement with previous works
in the literature as the ones of Unavane at al. (1996) and Venn et
al. (2004). On the other hand, as stated above, if we assume that the
Galactic halo formed by accreting enriched gas from dSphs or UfDs, we
also need a stellar contribution from dSphs and UfDs to explain the
stars at very low [Fe/H] that currently reside in the halo.

 It is worth noting that for the $\alpha$ elements studied in this
work, all models predictionsincluding the enriched infall of gas, tend
to the [$\alpha$/Fe] values of the reference model for high [Fe/H] in
the halo phase. This is due to the fact that when the SF is active the
pollution from dying stars overcomes the enriched infall effects.

 We note that a different value of the threshold in the MW
model would modify the duration of the phase 2). In this work, we did
not explore different values of the threshold in the halo, our aim being
to test the effects of the enriched infall on chemical evolution models
of the MW, which are able to reproduce the majority of the
observations in the solar neighborhood and also the abundance
gradients along the disk. As shown in Mott et al. (2013), if we do not
take into account radial gas flows (Portinari \& Chiosi 2000,
Spitoni \& Matteucci 2013, Cavichia et al. 2014) a threshold in the
gas density is required to explain the abundance gradients along the
Galactic disk, and in particular the values of 4
$\mathrm{M}_{\odot}\,\mathrm{pc}^{-2}$ in the halo phase and 7
$\mathrm{M}_{\odot}\,\mathrm{pc}^{-2}$ in the thin disk phase, provide
a very good agreement with the data.

In Fig. \ref{Ba1}, we show the results for the [Ba/Fe] versus [Fe/H]
abundance diagram.  The observational data are from  Frebel et
al. (2010), as selected and binned by Cescutti et al. (2013).  By
looking at the figure,  the 2IM model  does not provide a good
agreement with the observed data-set for
$\mathrm{[Fe/H]}<-2.5\,\mathrm{dex}$.  The initial increasing trend of
the [Ba/Fe] ratios in the 2IM model is due to the contribution of the
first Ba-producers, which are massive stars with mass in the range
$12$-$30\,\mathrm{M}_{\odot}$.

By looking at Fig. \ref{Ba1}, one can also appreciate that our
chemical evolution models for dSphs and UfDs fail in reproducing the
observed data, since they predict the [Ba/Fe] ratios to increase at
much lower [Fe/H] abundances than the observed data.  Concerning the
chemical evolution of the Ba for dSphs, our predictions are in
agreement with Lanfranchi et al. (2008), where they compared the
evolution of $s$- and $r$- process elements in our Galaxy  with 
that in dSph
galaxies.  That is due to the very low SFEs assumed for dSphs and
UfDs, which cause the first Ba-polluters to enrich the ISM at
extremely low [Fe/H] abundances. The subsequent decrease of the
[Ba/Fe] ratios is due to the large iron content deposited by Type Ia
SNe in the ISM, which happens at still very low [Fe/H] abundances in
dSphs and UfDs. Hence, in the range $-3.5\la \mathrm{[Fe/H]} \la
-2.5\,\mathrm{dex}$, while Galactic halo stars exhibit an increasing
trend of the [Ba/Fe] versus [Fe/H] abundance ratio pattern, UfD stars
show a decreasing trend (see also Koch et al. 2013).

\begin{figure*}
	  \centering \includegraphics[scale=0.3]{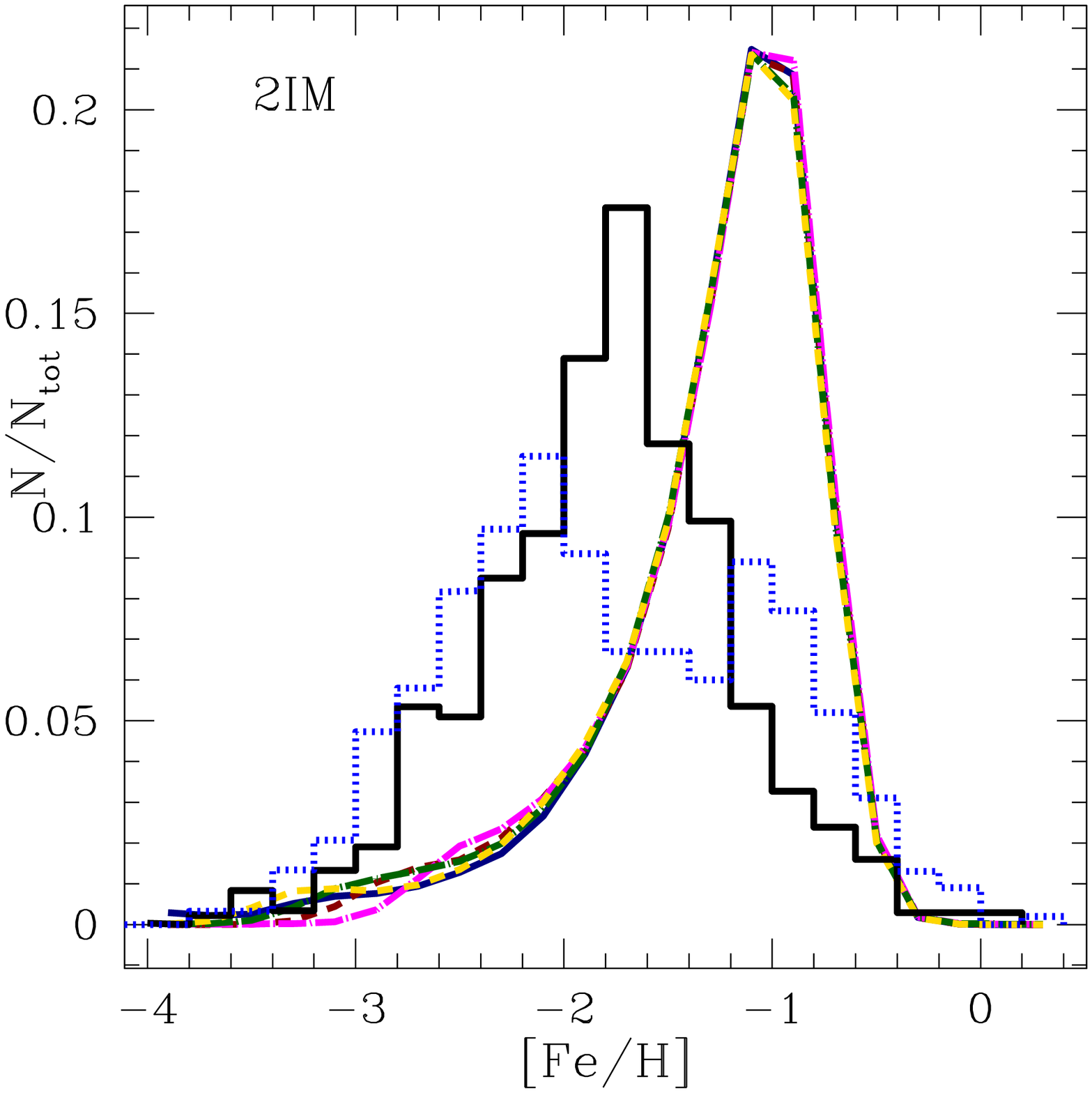} 
      \centering \includegraphics[scale=0.3]{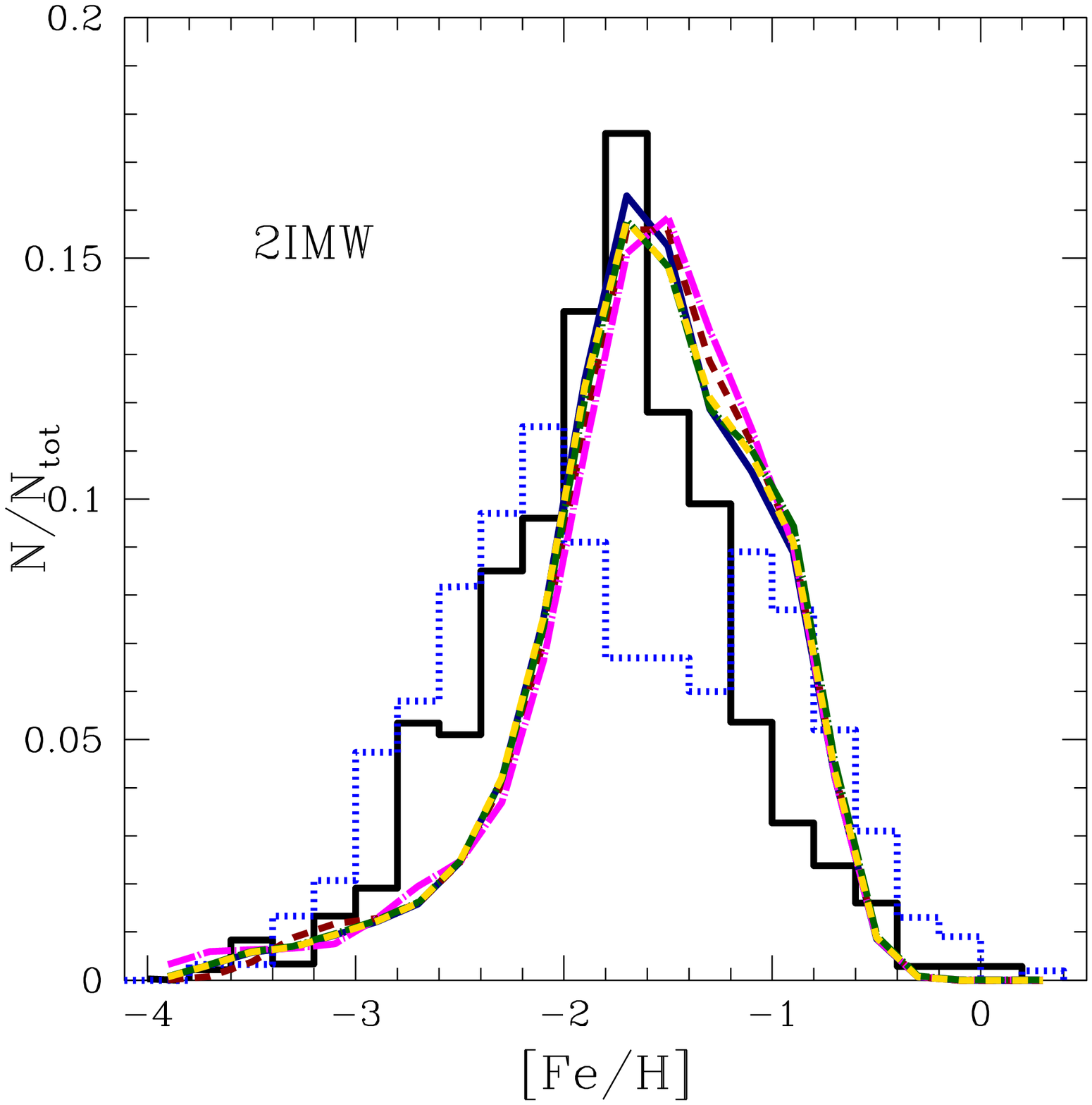} 
 \caption{ The G-dwarf metallicity distributions ([Fe/H]) predicted by  models with enriched infall of gas based on the reference 2IM model
({\it left panel}) and the ones based on the 2IMW model ({\it right panel}), are compared to the observed distributions by  Ryan \& Norris (1991, dotted blue histogram) and Sch\"{o}rck et al. 
(2009; solid black histogram). Concerning the {\it left panel},  model color lines are the same as in Fig. \ref{O1}, on the other hand in the {\it right panel} color lines are the same as in Fig. \ref{OW}. Our
predictions have been convolved with a Gaussian with an error of 0.2 dex. } 
\label{GAUSSFEH}
\end{figure*}

\begin{figure*}
	  \centering \includegraphics[scale=0.3]{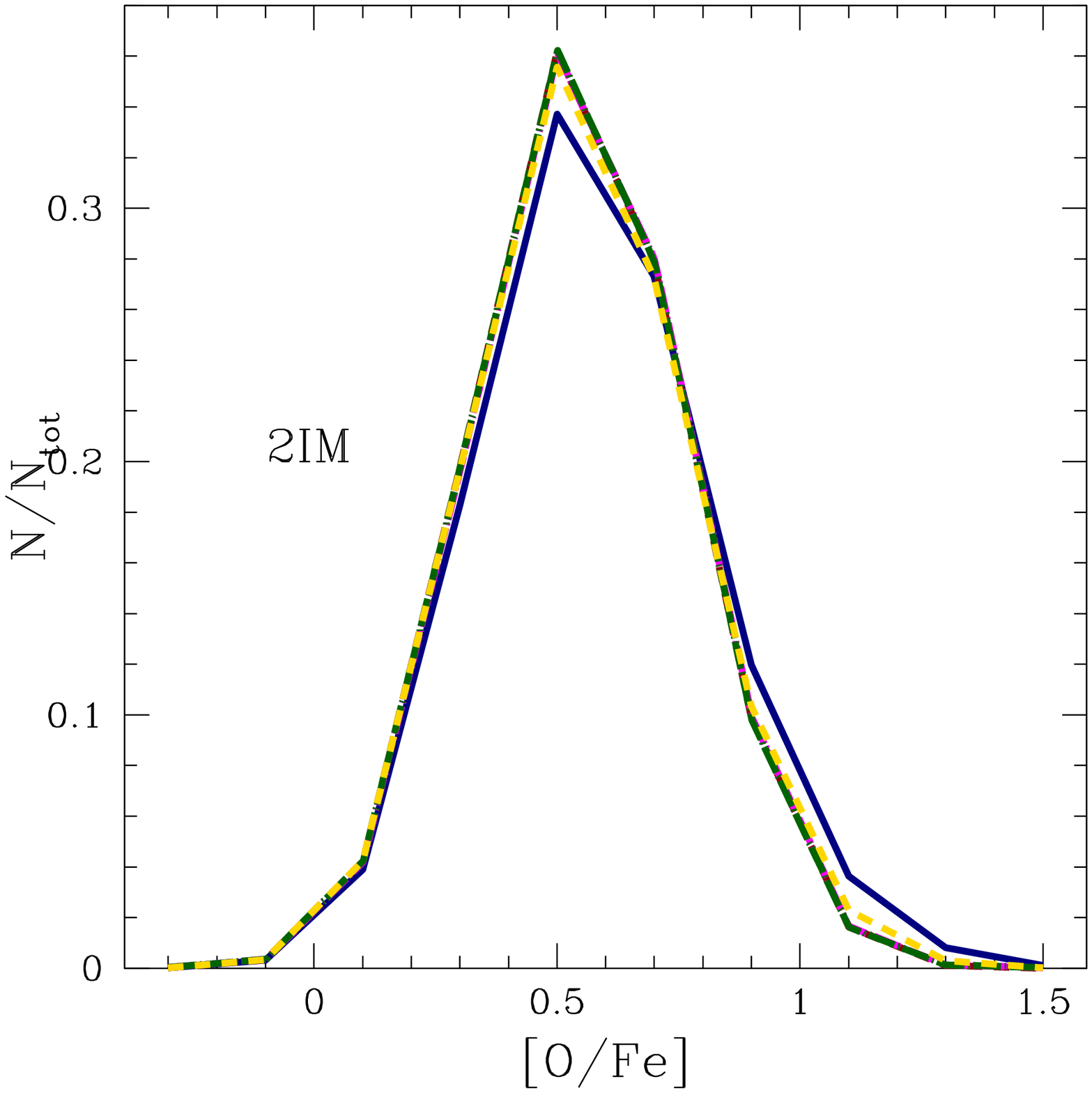} 
 \centering \includegraphics[scale=0.3]{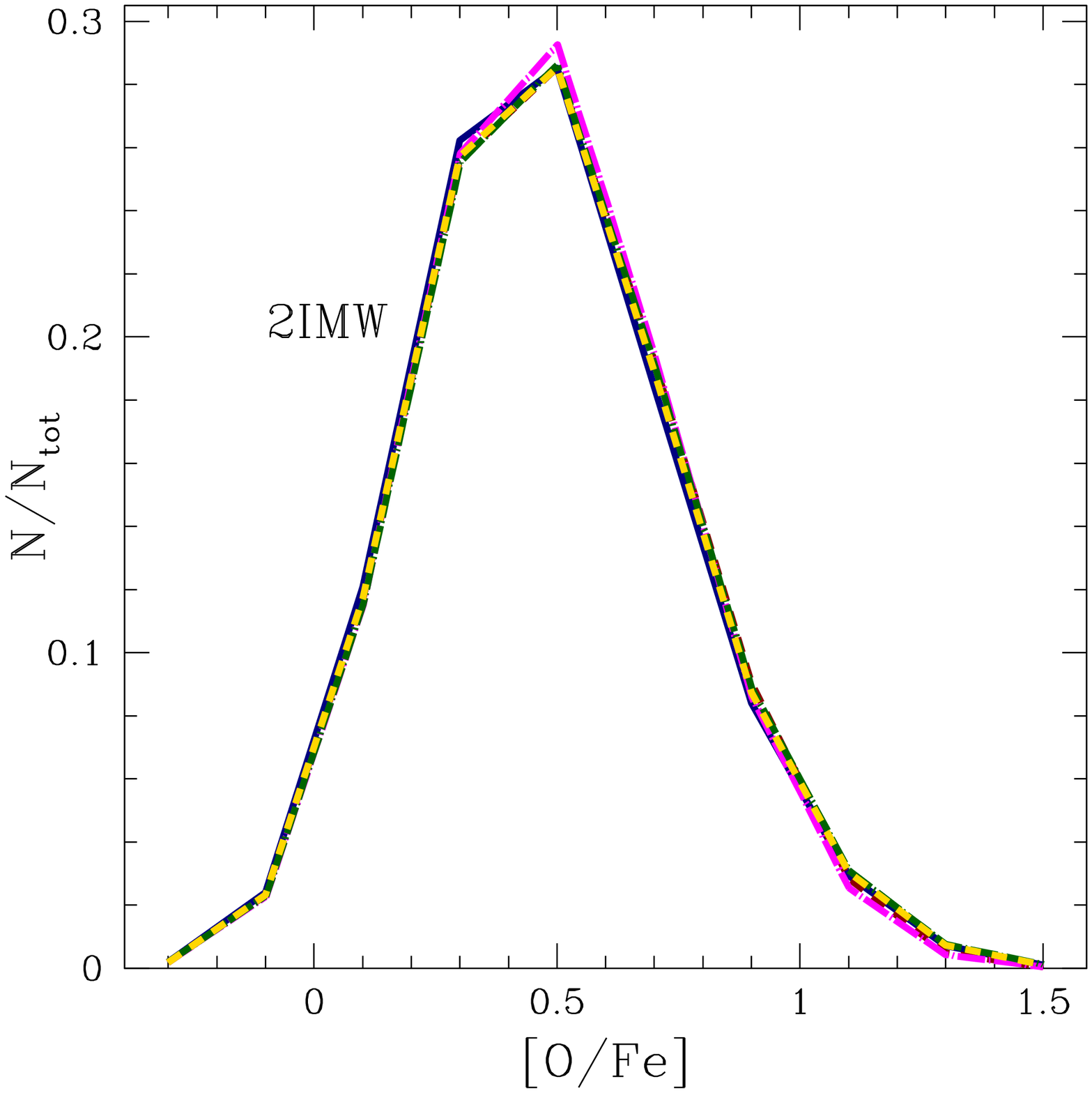} 
      \caption{  The G-dwarf  distributions in terms of [O/Fe] predicted by  models with enriched infall of gas based on the reference 2IM model
({\it left panel}) and the ones based on the 2IMW model ({\it right panel}). Concerning the {\it left panel},  model color lines are the same as in Fig. \ref{O1}, on the other hand in the {\it right panel} color lines are the same as in Fig. \ref{OW}. Our
predictions have been convolved with a Gaussian with an error of 0.2 dex. } 
\label{GAUSSOFE}
\end{figure*}

In Fig. \ref{Ba1}, all our models involving an enriched infall from
dSphs and UfDs deviate substantially from the observed trend of the
[Ba/Fe] versus [Fe/H] abundance pattern in Galactic halo stars. Such a
discrepancy enlarges for $\mathrm{[Fe/H]}<-2.4$ dex, where those
models predict always larger [Ba/Fe] ratios than the 2IM model.
 
\subsection{The Results: the Galactic halo in the model 2IMW}

In this subsection we show the results when the time dependent
enriched infall is applied to the reference model 2IMW. In
Fig. \ref{OW} we show the results in terms of [O/Fe] versus [Fe/H] in the
solar neighborhood.

As mentioned in Section 2, the reference model 2IMW in the halo phase
is characterized by an outflow gas rate  proportional to the SFR
and by a smaller formation timescale $\tau_{H}$ than the one of the model
2IM (see Table 1). A shorter formation time-scale leads to a faster
chemical evolution at the early times. In fact, as shown in Fig. 2 of
Brusadin et al. (2013), the SFR starts at 0.05 Gyr, since the critical
threshold in the surface gas density is reached earlier.

As done above,  for the  model ``2IMW+dSph'' we  distinguish three different phases in the halo chemical evolution:
\begin{itemize}

\item Phase 1): 0-0.013 Gyr, the infall is primordial, the wind in dSphs has not started yet {and there is no star formation};
\item Phase 2): 0.013-0.05 Gyr, the infall is enriched by dSphs, the SFR is zero in this phase;

\item Phase 3): 0.05- 1 Gyr; the infall is enriched by dSphs, the SFR is different from zero.
\end{itemize}

On one hand, comparing  model ``2IMW+dSph'' in Fig. \ref{OW} with  model ``2IM+dSph''
in Fig. \ref{O1}, we can see that the former  shows a   shorter
phase 2) than the latter.

On the other hand, comparing model ``2IMW+UfD'' in Fig. \ref{OW} with
model ``2IM+UfD'' in Fig. \ref{O1}, we see that the former overlap to
the reference model 2IMW at almost all [Fe/H] abundances.  In fact,
since in the UfD galactic model the wind starts at 0.088 Gyr and, at
this instant, in the model 2IMW the SF is already active (the SF
activity begins at 0.05 Gyr), the phase 2) evolution (namely, chemical
evolution without SF) is not present in the 2IMW+UfD model.  Therefore
the effect of the enriched infall is almost negligible compared to the
pollution of chemical elements produced by dying halo stars, since we cannot
distinguish between the effect of an enriched infall from UfDs and the
chemical feedback provided by the Type II SNe originated in the
Galactic halo itself.

This can also be appreciated by looking at Fig. \ref{OHtW}, where the time evolution of the oxygen abundances  of the model 2IMW, and in
presence of enriched infall from UfD and dSph galaxies are shown. We
see that for the UfD case the model with enriched infall is almost
identical to the reference model 2IMW.

 In Figs. \ref{siw} and \ref{mgw} model results for  [Si/Fe] vs. [Fe/H] and   [Mg/Fe] vs [Fe/H] in the solar neighborhood are presented, respectively. As for oxygen the effect of the enriched infall from UfD galaxies is almost negligible compared to the
pollution of chemical elements produced by dying halo stars.    

Concerning the [Ba/Fe] versus [Fe/H] abundance pattern, in Fig. \ref{BaW}
we compare the predictions of our models with the Galactic halo data.
We notice that the 2IMW model provides now a better agreement with the
observed data than the 2IM model, although the predicted [Ba/Fe]
ratios at $\mathrm{[Fe/H]}<-3\,\mathrm{dex}$ still lie below the
observed data.  On the other hand, by assuming an enriched infall from
dSph or UfD galaxies, the predicted [Ba/Fe] ratios agree with the
observed data also at $\mathrm{[Fe/H]}<-3\,\mathrm{dex}$.  In
conclusion, in order to reproduce the observed [Ba/Fe] ratios over the
entire range of [Fe/H] abundances, a time-dependent enriched infall in
the Galactic halo phase is required.  We are aware that  for Ba  more detailed data are needed, therefore at this stage we cannot draw firm
conclusions.

We note that in general [$\alpha$/Fe] ratios in dSphs and UfDs
can overlap with those of halo stars at very low metallicity where in
all galaxies the chemical enrichment is dominated by the
nucleosynthesis of core collapse SNe. 

This makes difficult to test whether some stars from dwarf satellites
have indeed been accreted by the halo.  Rather, it is likely that a
fraction as high as 50\% of the gas out of which the halo formed has
been shed by its satellite systems, whose relics we see (in part)
nowadays, devoid of their gas.  This is only valid for satellite
systems with a very short duration of star formation, as modelled in
this paper.  More realistic SFH would create ejected gas with low
[$\alpha$/Fe] (as seen in the stars in the surviving luminous dSph)
and conflict with the data for halo stars.

On the other hand, we have
identified in the [Ba/Fe] ratios a better discriminant of the origin
of halo stars. The differences predicted for the dwarf galaxies
relative to the halo suggest that it is unlikely that the dSphs and
UfDs have been the building blocks of the halo.

 In Fig. \ref{GAUSSFEH} we show the predicted G-dwarf distributions in terms
of [Fe/H] for models of Figs. 3 and 8.  Our
predictions have been convolved with a Gaussian with an error of 0.2 dex. As pointed out by Brusadin
et al. (2013),  the reference model 2IM is not able to reproduce the
peak of the distribution. Only assuming a shorter formation time-scale
coupled with a gas  outflow event in the Galactic halo chemical evolution
 (model 2IMW) we are able to properly fit the observed
distribution. Anyway, in both cases the enriched infall of gas from dSph and UfD objects  does not affect the distributions. In Fig. \ref{GAUSSOFE} the G-dwarf distribution is shown as a function of the [O/Fe] ratio. Here, both models 2IM and 2IMW, including the enriched infall models show the peak at 0.5 dex. 

\section{Conclusions}

In this paper,  we  first have explored the hypothesis that dSph and UfD
galaxies are the survived building blocks of the Galactic halo, 
  by assuming that the halo formed by accretion of stars belonging
to these galaxies.

  Then,  we have presented a different  scenario in which the Galactic halo
formed by accretion of  enriched gas  with  the same chemical   composition as the outflowing gas  from dSphs and
UfDs. Finally, we have tested the effect of diluting the infalling
material from dSphs or UfDs with primordial gas of different
extra-galactic origin.

Our main conclusions can be summarized as follows:

\begin{enumerate}

\item We find that the predicted   [$\alpha$/Fe] versus [Fe/H] abundance patterns of UfD and dSph chemical evolution models deviate substantially from   the observed data of the Galactic halo stars only for [Fe/H] values larger than -2 dex;  this means that at those metallicities  the chemical evolution of the Galactic halo was different than in the satellite galaxies. On the other hand,  we notice that for Ba the
chemical evolution models of dSphs and UfDs fail to reproduce the
observational observed data of the Galactic halo stars over the whole
range of [Fe/H].

\item We can safely rule out the hypothesis that  
the stellar halo of the MW entirely formed from the merging of
galaxies which were the ancestors of the current dSphs and UfDs.  Our
results are in agreement with the previous suggestions of Unavane et
al. (1996) and Venn et al. (2004).  We cannot rule out, however, the
hypothesis that a substantial contribution to the formation of the
Galaxy stellar halo was provided by a population of dwarf galaxies
which were more massive and more evolved from the point of view of the
ISM chemical evolution than the current dSphs and UfDs.

\item Concerning the chemical evolution models for the MW in the presence of  enriched gas infall we obtain that: the effects  on the [$\alpha$/Fe] versus [Fe/H] plots depend on the infall time scale for the formation of the halo and the presence of a gas threshold in the star formation. In fact, the most evident effects are present for the model 2IM, characterized by  the longest time scale of formation (0.8 Gyr), and  the  longest period without star formation activity  among  all models  presented here.

\item In general,  the enriched infall by itself is not capable to explain the  observational spread in the halo data at low  [Fe/H], in the  [$\alpha$/Fe] versus [Fe/H] plots. Moreover, in the presence of an enriched infall  we need  stars produced in situ in dSph or UfD objects  and accreted later to the Galactic halo, to explain the data at lowest [Fe/H] values.

\item The optimal element to test different theories of halo formation is barium which  is (relatively) easily measured in low-metallicity stars. In fact, we  have shown  that  the predicted  [Ba/Fe] versus [Fe/H] relation  in dSphs and UfDs is quite different than in the Galactic halo.
Moreover, the [Ba/Fe] ratio can be  
substantially influenced by the assumption of an enriched infall.  In
particular, the two infall plus outflow model can better reproduce the
data in the whole range of [Fe/H] abundances, and this is especially
true if a time dependent enriched infall during the halo phase is
assumed.

\end{enumerate}

\section*{Acknowledgments}
We thank the anonymous referee for the suggestions that
improved the paper.
We thank L. Gioannini for many useful discussions.  The work was
  supported by PRIN MIUR 2010-2011, project ``The Chemical and
  dynamical Evolution of the Milky Way and Local Group Galaxies'',
  prot. 2010LY5N2T.

\end{document}